\documentclass[aps,prl,twocolumn]{revtex4-1}
\usepackage{amsmath}
\usepackage{amssymb}
\usepackage{graphics}
\usepackage{epsfig}
\usepackage{color}
\usepackage{graphicx}
\usepackage{color}
\begin{document}
\title{Emergence of diversity in a model ecosystem}
\author{Namiko Mitarai, Joachim Mathiesen, Kim Sneppen}

\affiliation{Niels Bohr Institute,  
University of Copenhagen, Blegdamsvej 17, DK-2100 Copenhagen, Denmark}

\begin{abstract}
The biological requirements for an ecosystem to develop and maintain species diversity are in general unknown. Here we consider a model ecosystem of sessile and mutually excluding organisms competing for space [Mathiesen et al. Phys. Rev. Lett. {\bf 107}, 188101 (2011)]. Competition is controlled by an interaction network with fixed links chosen by a Bernoulli process.  New species are introduced in the system at a predefined rate. In the limit of small introduction rates, the system becomes bistable and can undergo a phase transition from a state of low diversity to high diversity. We suggest that isolated patches of meta-populations formed by the collapse of cyclic relations are essential for the transition to the state of high diversity.
\end{abstract}
\maketitle

\section{Introduction}
Multiple species often coexist robustly in natural ecosystems\cite{tilman,ives}. Species interact with each other \cite{eigen,tejedor,smith} primarily through competition, cooperation or predation \cite{lotka,volterra,may,dayton,paine}. It is however not easy to keep an ecosystem with diversity when several species compete for the same resources \cite{gause,hardin}. It has been found that one way to maintain coexistence of multiple species is to include hypercycles or predator-prey cycles. Especially the cycle of three species has been studied in detail. Coexistence of oscillating populations of three species is found to be stable in the deterministic case, while noise due to a finite number of agents always leads the system to a single species state \cite{Reichenbach2006}. Another robust way to maintain high diversity is to include space \cite{mollison,hufkens,kareiva,kerr,schrag,heilmann}.  Combinations of space and cycles \cite{bjoerlist, gilpin,Perc2007, Laird2006,Szabo2001} are found to maintain stable coexistence. Further, non-transitive allelopathic relationships with competition in 2-dimensional space  \cite{jackson,buss,karlson} have been found to be able to maintain species diversity on a longer timescale than pure hierarchical predation relationships.

In our previous letter \cite{Mathiesen2011},  a model ecosystem of sessile and mutually excluding organisms has been introduced, where competition for resources is a zero-sum game about available space. The model considers predation or allelopathic interaction between species, where a species can invade a space that is already occupied by another species. It has been demonstrated that the model shows a sharp transition from multiple to single species as the number of ``predation'' interactions is increased. It has been suggested that
the important mechanism behind the increased diversity is
spatial fragmentation of populations creating isolated niches for new species.
In this paper, we present a detailed analysis of the model ecosystem focusing on the necessary mechanism for maintaining a high species diversity. We first demonstrate that the system shows clear bistability between the low- and high-diversity states, and the transition between the states is triggered by fragmentation of populations into many patches. We then demonstrate that cyclic relations of four and more species result in many isolated stable patches when one of the species spontaneously dies out due to noise. This noise is necessary in order to maintain high diversity.

The paper is divided in 4 sections. In Section 2, we introduce a model ecosystem where sessile species compete for space \cite{Mathiesen2011}. In Section 3, the creation of diversity in the ecosystem is analyzed, and finally, in Section 4, we provide a discussion of the stability of the model.

\section{Model ecosystem }
Our model is inspired by the spatial dynamics of lichen communities\cite{JNHTM10, Mathiesen2011}. When a crustose lichen meets another, a contact boundary is formed. The boundary remains stable over time if the species are competitively equal, but sometimes bulging boundaries between species can be observed, which suggests some species can take over another. The model might also provide insight into the evolution of seaweed \cite{Stebbing1973} or whole epifaunal communities \cite{KayandKeough1981}, which grow essentially in two dimensions and often lack one dominating species. In our simple ecosystem model, we consider an ecosystem of multiple-species competing on a two dimensional lattice. At any given time a lattice site can be occupied by one species only. The species on a site can invade a neighbor site, provided that it is occupied by a competitively inferior species.

We characterize the ecosystem by a directed network of possible species interactions. Interactions are materialized only when organisms connected by a link are neighbors somewhere in the system. The aim of our model is to study ecosystem diversity as we change the number of potential interactions between species, parametrized by $\gamma$. In addition to this, new species are introduced to the system at a rate $\alpha$.

Each (time) step of our model consists of two possible events:
(i)
Select a random site $i$ and one of its nearest
neighbors $j$. If the species $s(i)$ at site $i$ can invade the species
$s(j)$ at site $j$, i.e. $\Gamma (s(i),s(j))=1$, then site $j$ is
updated by setting $s(j)=s(i)$. Here  $\Gamma$ is the matrix that
represents the possible interactions.
(ii)
With probability $\alpha \times \gamma /N$ per site a new random species $s$ is introduced
at a random site $j$ and assigned random interactions $\Gamma (s,u)$
and $\Gamma (u,s)$ with
all existing species $u$ in the system.
Each of these interactions are
assigned value 1 with probability $\gamma$, or otherwise set to $0$ (we do allow for the case $\Gamma (u,s)=\Gamma (s,u)=1$).
The introduced species $s$ is assumed to be able to invade the previous species
at the site $j$, $s(j)$: $\Gamma (s,s(j))=1$. (This is the reason why the introduction probability is proportional to $\gamma$; this is equivalent to the procedure where a new species $s$ tries to invade a randomly selected site $j$ at the rate $\alpha$ per system but it succeeds only if  $\Gamma(s,s(j))=1$.  )
$\Gamma (u,s)$ and $\Gamma (s,u)$ will not change
once they are introduced.

One time unit is defined as $N$ repeats of procedure (i) and (ii),
which means, per time unit, 
on average each site makes one attempt
to invade a neighbor, { and $\alpha$ new species attempt to enter 
the system (the number of new species appearing
is $\alpha \times \gamma$ on average per unit time)}.

{ In the following simulations and unless otherwise noted, we initialize the { empty} system by introducing a new species for every step during the first 100 time steps. After a short transient this initialization leads to a system with a dynamics fully equivalent to that of a system started with just one species.
}

{Note that this model allows two types of
``competitively equal'' species pairs, i.e.,
$\Gamma(u,v)=\Gamma(u,v)=0$ or $\Gamma(u,v)=\Gamma(u,v)=1$.
The former case represents a situation where
$u$ and $v$ can grow equally well, but one cannot
take over the region that is already occupied by the other,
hence the boundary between $u$ and $v$ is stable.
The latter case represents a more ``aggressive'' situation,
for example the species $u$ produces toxin to kill the species
$v$ and vise versa, resulting in fast and noisy modification of the
boundary between the mutually aggressive species.}

\begin{figure}
\includegraphics[width=0.4\textwidth]{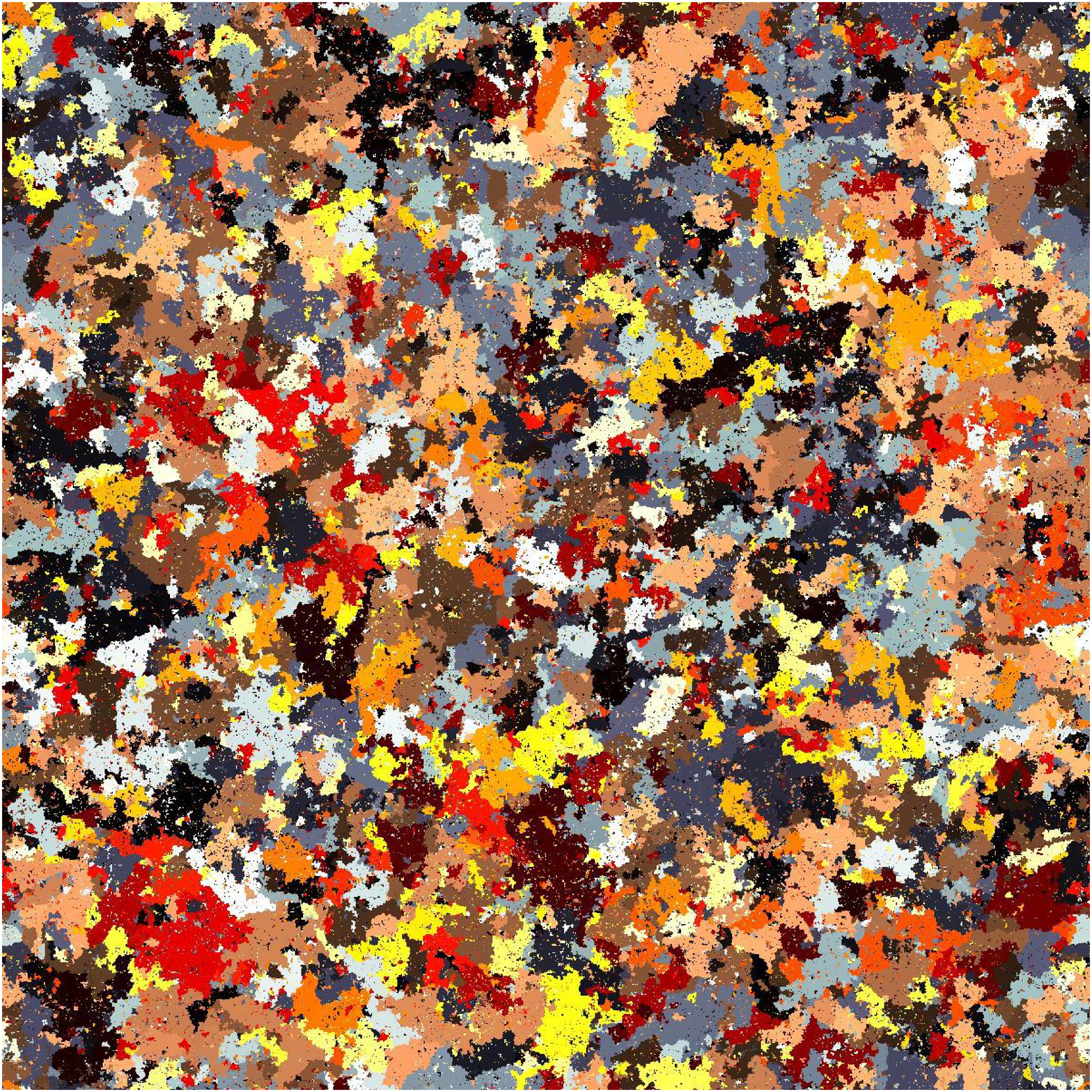}
\includegraphics[width=0.45\textwidth]{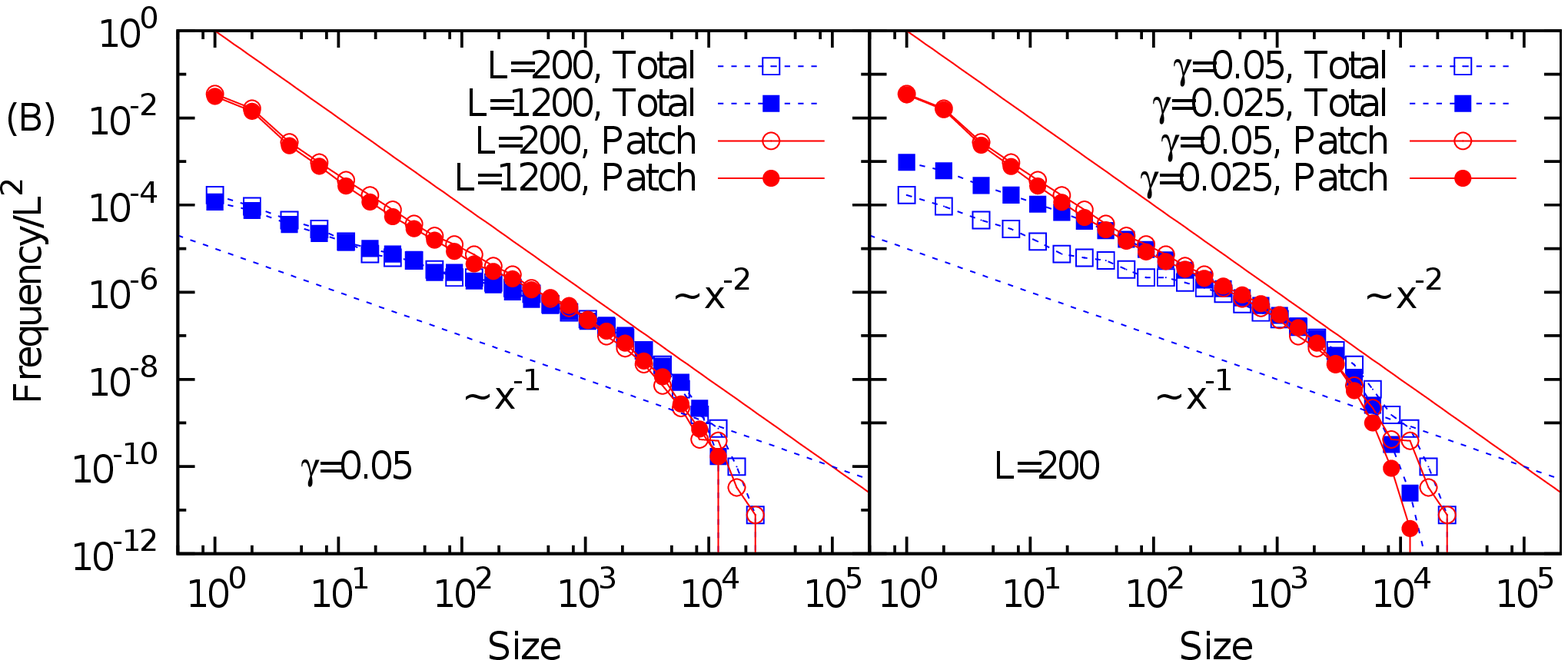}
\caption{(color online) (A) Figure of a system of size $L=1200$ with $\gamma=0.05$ and  $\alpha=0.01$. (B) Distributions for the total species sizes (squares) and the patch sizes (circles) for different $L$ (left panel, $\gamma=0.05$) and different $\gamma$ (right panel, $L=200$) { in the steady state.
$\alpha=0.01$ for $L=1200$, and $\alpha=0.0025$ for $L=200$.
The data for $L=200$ are averaged over $4\times 10^8$ time steps.} 
 The vertical axis shows the probability scaled by the system size $L^2$ of finding a species or patch of given size.
}\label{1200sys}
\end{figure}

\section{Results}
\subsection{Phase transition between low- and high-diversity state }
Previously \cite{Mathiesen2011}, we have found that the model shows, as the interaction probability $\gamma$ is decreased from one,  a first-order phase transition from a low-diversity to a high-diversity state at a critical value of $\gamma=\gamma_c \approx 0.055$ in the limit of $\alpha \to 0$. Remarkably, the high diversity state does not exist in the random-neighbor { variant} of the model: When each site is allowed to access every other site in the system, { the interaction matrix $\Gamma$ is not enough to keep high diversity even for
small $\gamma$, and} the diversity $D$  (defined as number of species in the system)  always approaches 1 for small $\alpha$  \cite{Mathiesen2011}.

\begin{figure}
\includegraphics[width=0.45\textwidth]{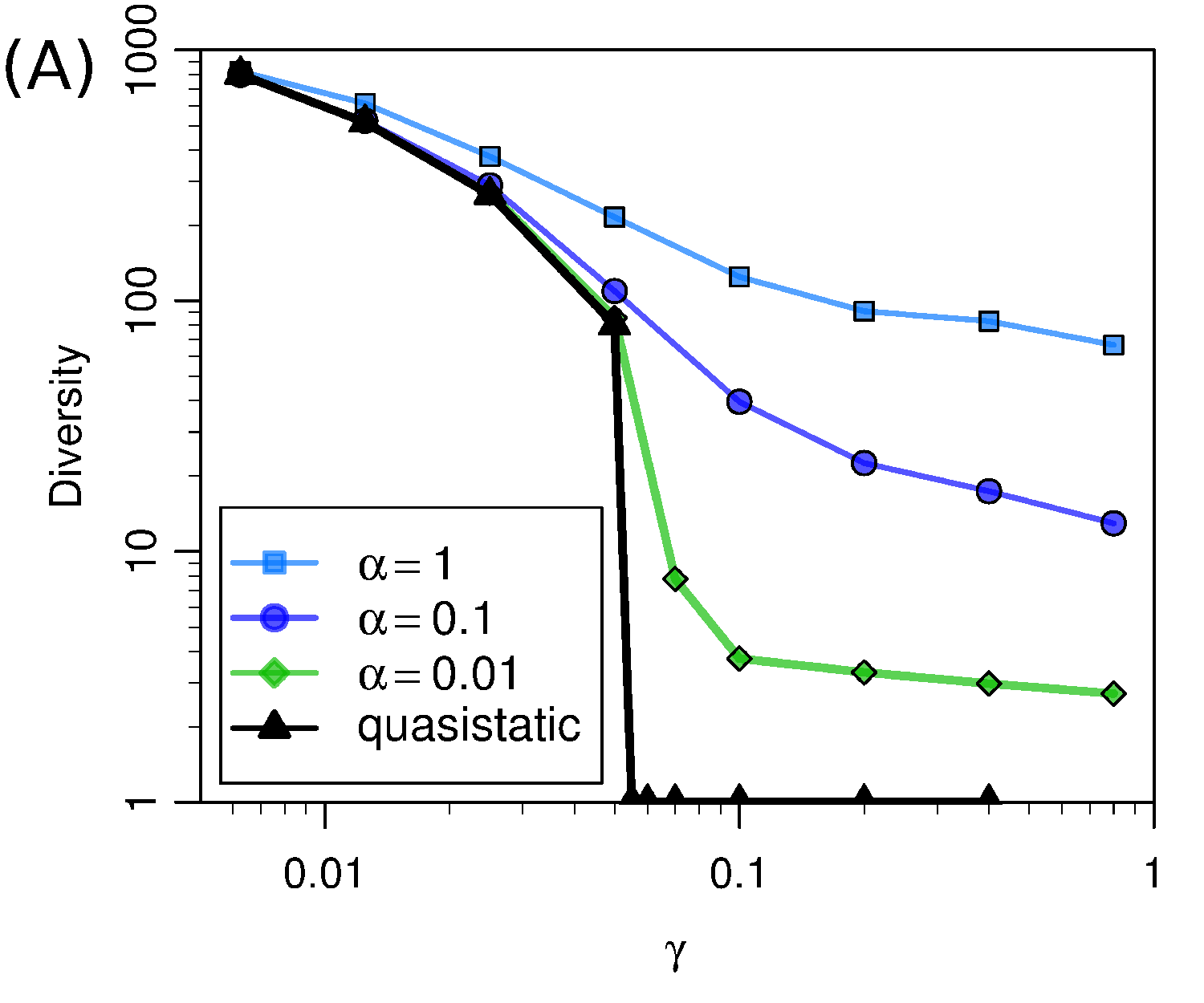}
\includegraphics[width=0.45\textwidth]{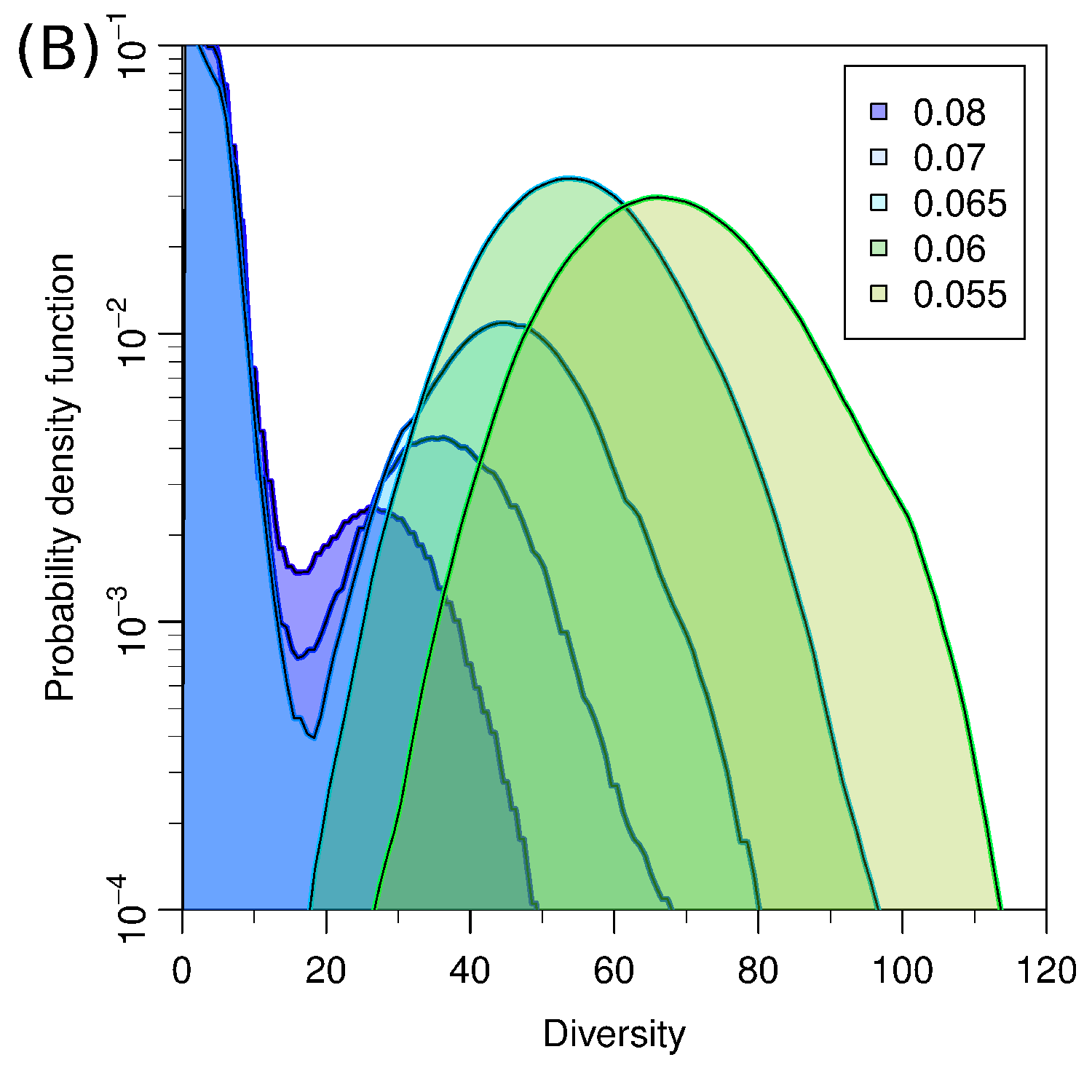}
\caption{(color online) (A)
The time-averaged diversity $D$ as a function of $\gamma$ for various $\alpha$.
A sharp transition is seen in the limit of $\alpha\to 0$. (This corresponds to
quasistatic simulation, explained later in the text.) { The connecting lines are for guiding the eye.}
(B) Probability density function of the diversity measured over time for a system of size $L=200$. The distributions are shown for a few values of $\gamma$ below and above the critical value $\gamma_c$. We see that for $\gamma\geq .06$ the distributions are bimodular. The density is shown along a logarithmic axis.}
\label{phasediagram}
\end{figure}

The high diversity state is illustrated in Figure ~\ref{1200sys} (A), which presents a snapshot  of a model ecosystem of size $N=L\times L= 1200\times 1200$, with $\gamma=0.05$ and $\alpha=0.01$. The different colors represent different species.  We can see that the species are fragmented into many patches and that there is no cluster spanning the whole system.
{ In contrast, when increasing $\gamma$ to 0.065,
the high-diversity state is only metastable and
collapses to the low species state,
where one dominating species interspersed with a few small patches
of other species (see also Fig. \ref{phasediagram}AB).}

The transition between a high and low diversity state can be obtained only for large enough system size ($L\ge 200$).  { As long as the system size is large enough,} the diversity $D$ was found to follow a simple scaling of $D\propto L^2$ in the high diversity regime \cite{Mathiesen2011}. This can be illustrated more clearly by analyzing snapshots of high diversity states.

Figure ~\ref{1200sys} (B) shows the distribution of species size or abundance (number of sites occupied by a species) and patch size (number of spatially connected sites occupied by the same species), where different system sizes are compared in the left panel. We can see that both distributions change quite little between $L=200$ and $L=1200$, and show power law behavior with exponent $-1$ ($-2$) for species abundance (patch size) in the small area regime, while there is a clear cut-off around 5,000.  This observation agrees with the scaling of $D\propto L^2$ for systems large enough compared to the cut-off.

The high abundance of small patches compared to rare species implies that part of the species are fragmented into many small patches (meta-populations). It is this fragmentation that results in a sharper exponent for the patch size distribution than the species abundance distribution.  When $\gamma$ is decreased (Fig.~\ref{1200sys} (B) right),  the diversity $D$ increases, but surprisingly the patch size distribution remains unchanged. Hence, the species-abundance distribution must have more weight around the small abundances. Namely, the increase of diversity with smaller $\gamma$ is simply due to the replacement of species in existing patches.

The first-order transition obtained when increasing $\gamma$ above the critical $\gamma_c\approx 0.055$ is depicted in
Fig. \ref{phasediagram}(A), where the time-averaged diversity $D$ is
shown for various values of $\alpha$.
It shows a sharp transition in the limit of $\alpha\to 0$.
This transition is associated with bistability of the overall system behavior, as depicted in Fig.~\ref{phasediagram}(B).  This figure shows the probability distribution of $D$ for various $\gamma$  values, for system size $L=200$ and $\alpha=0.01$. We can see that, above  $\gamma_c$ the probability distribution for the diversity has two peaks at low and high diversity, respectively. In contrast, for $\gamma$ below $\gamma_c$, the high-diversity state becomes monostable. 

\subsection{Emergence and collapse of diversity}
\begin{figure}
\includegraphics[width=0.45\textwidth]{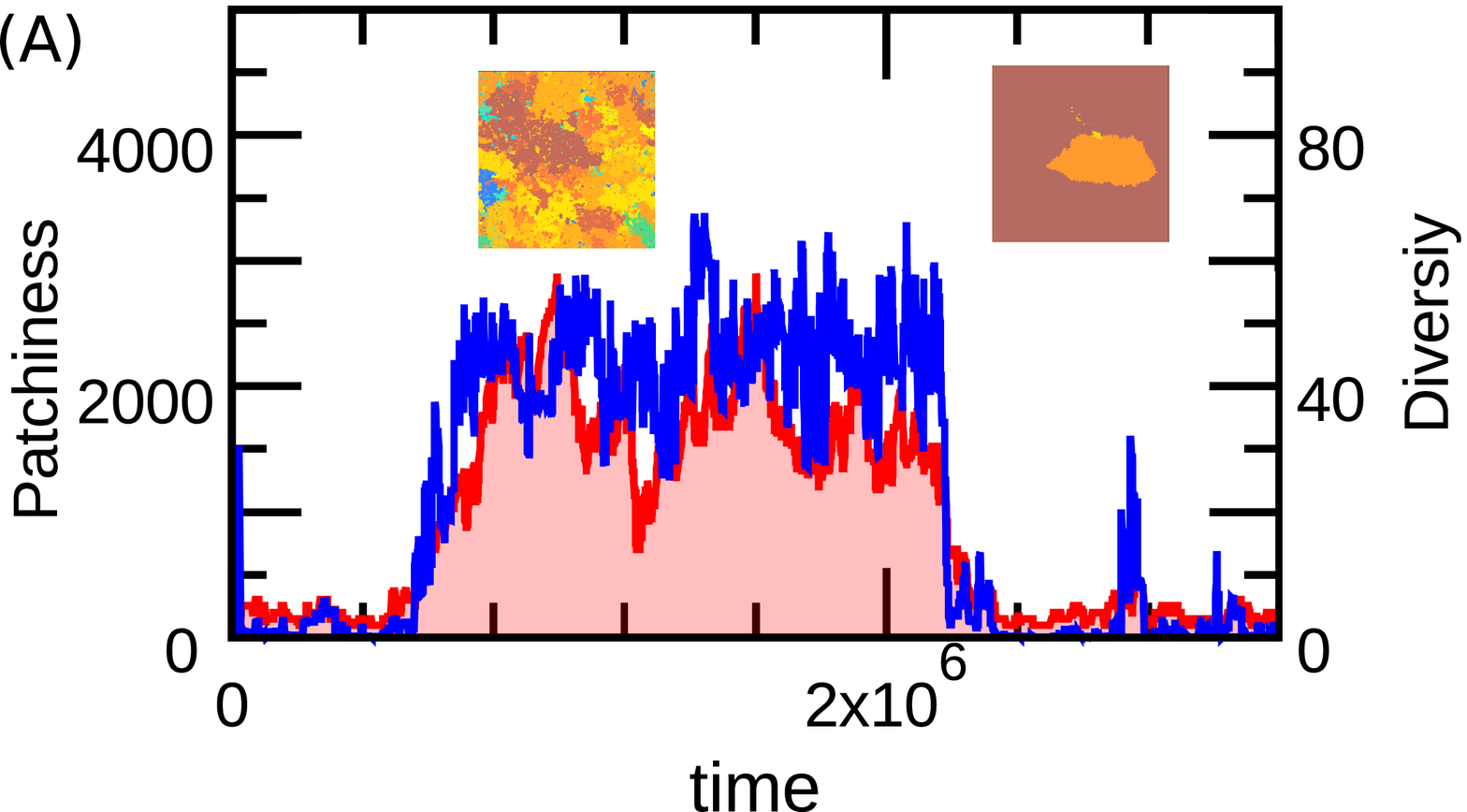}
\includegraphics[width=0.45\textwidth]{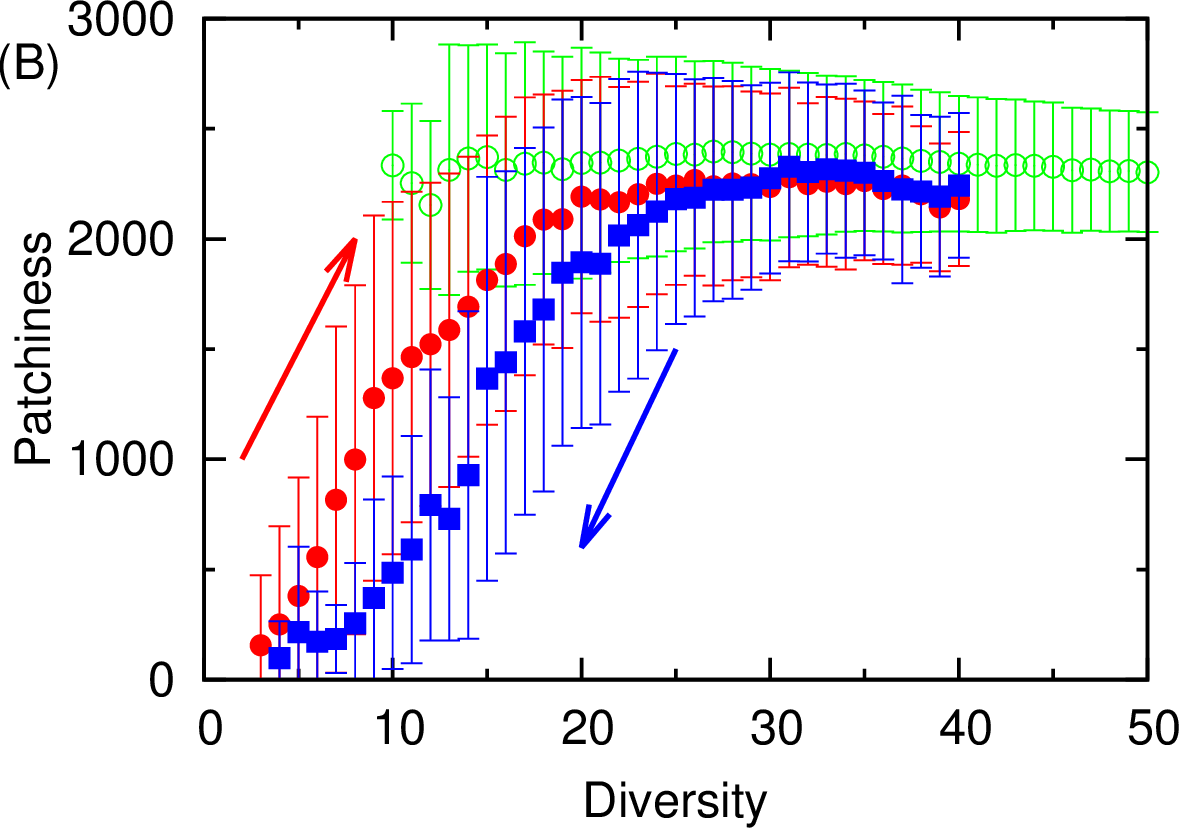}
\caption{(Color online) (A) Time evolution of the patchiness (line) and diversity (line with shaded area) for { $\alpha=0.0125$} and $\gamma=0.07$ for a system of size $L=200$.  {For illustration, typical snapshots from another simulation data set with the same parameters  are shown for low- and high-diversity states.}
(B) Relation between the patchiness and  diversity for { $\alpha=0.0125$}  and $\gamma=0.07$, averaged over 100 transitions between low and high diversity. The transition from low to high (filled circles) diversity is defined as a development where diversity increases from 3 to above 40. The reverse transition is marked by filled squares, and defined as a development where the diversity decreases from 40 to below 3. The patchiness is averaged for each values of diversity $D$, and the error bars show the standard deviations.  
As marked by the open circles, the patchiness is independent of the diversity as long as the system is staying in the high-diversity state. }\label{PatchTimeSeries}
\end{figure}

Next, we analyze the dynamics of the transition between high- and low-diversity states in the bistable parameter regime. Figure \ref{PatchTimeSeries}(A) shows the time evolution of the patchiness $P$ (number of patches in the system) and the diversity $D$, for { $\alpha=0.0125$}, $\gamma=0.07$, and $L=200$. We observe that the transition from low- to high-diversity occurs at time $6 \times 10^5$, and from high- to low-diversity at { time $2.2 \times 10^6$}. The transition looks rather sudden on this scale, but in fact  it develops over thousands of time units. 
In addition, the transition from high to low diversity is not simply a reversal of the transition from low to high. Rather, in both directions, the change in diversity is preceded by a change in patch size.

This tendency is confirmed in Fig.~\ref{PatchTimeSeries}(B), which shows average patchiness vs. diversity over ~100 transitions in both directions.
Figure~\ref{PatchTimeSeries}(B) also shows that, when the system is in the high diversity state (open circles), the patchiness stays at a high constant value, even though the diversity fluctuates between around 10 and 60. In contrast, when the system is at low diversity, the patchiness-diversity plot follows the branch of the high- to low-diversity transition (data not shown). Therefore a substantial increase of patchiness is the necessary condition for reaching the stable high diversity state.

\subsection{Stochastic patch creation}

Emergence of spatial fragmentation through patch creation is essential for
creation of biodiversity in our model. There are, however, two fundamentally different ways to create patches, both using the stochastic fragmentation occurring when one or more species are driven to local extinction.

\begin{figure}
\includegraphics[width=0.4\textwidth]{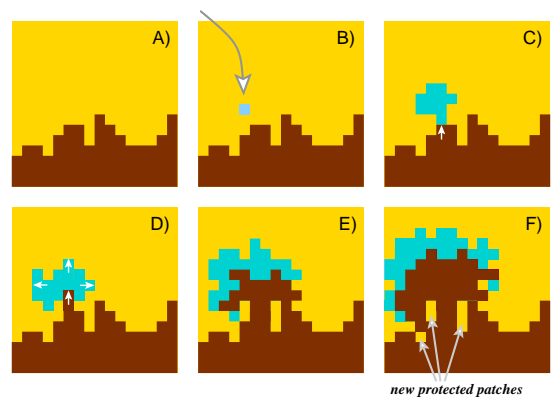}
\caption{(Color online) Stochastic patch creation in a system when a new species ``light blue (middle darkness)'' can invade yellow (lightest), but at the same time ``light blue (middle darkness)'' can be invaded by one of the species (``Brown'', darkest). The figure illustrate that patches can be created when species form this purely hierarchical relationships in a noisy system.
}\label{patch1}
\end{figure}

First we show in Fig.~\ref{patch1} that patches can be created by a purely linear relationship with, for example, three species A $\rightarrow$ B $\rightarrow$ C. While A and C alone can coexist stably with a well defined fixed border, the introduction of B makes it possible for A to invade the territory of C. First, B could, if undisturbed, completely eliminate C. However, when scavenging the borders of the region occupied by C,  B encounters A, which can invade B. A will then immediately start invading the area of B in parallel with B's invasion of C. As the dynamics is noisy, occasional patches of C will be left behind within the region of species A, because A cannot invade C directly.

A second mechanism for patch creation is associated with cyclic relationships. ``Active'' cycles are known to produce many fragmented areas, due to a continuous activity of invasions. However, this activity collapses if one of the species goes extinct. In the case of a three-cycle (A$\to$B$\to$C$\to$A, ‚ ``rock-paper-scissors''), the collapse leads to a homogeneous state with a single species (e.g., if A dies out, B will fully invade the area covered by C). However, if the cycle length is greater than or equals to four, there is a possibility that a finite number of patches are left even after the dynamics spontaneously has come to a stop due to the stochastic extinction of one of the species (e.g. in a cycle of A$\to$B$\to$C$\to$D$\to$A, if A dies out and B displaces C before C can displace D, then B and D will coexist). Thereby, a large enough cycle can create stationary patches without requiring long-lasting dynamics.

\begin{figure}
\includegraphics[width=0.45\textwidth]{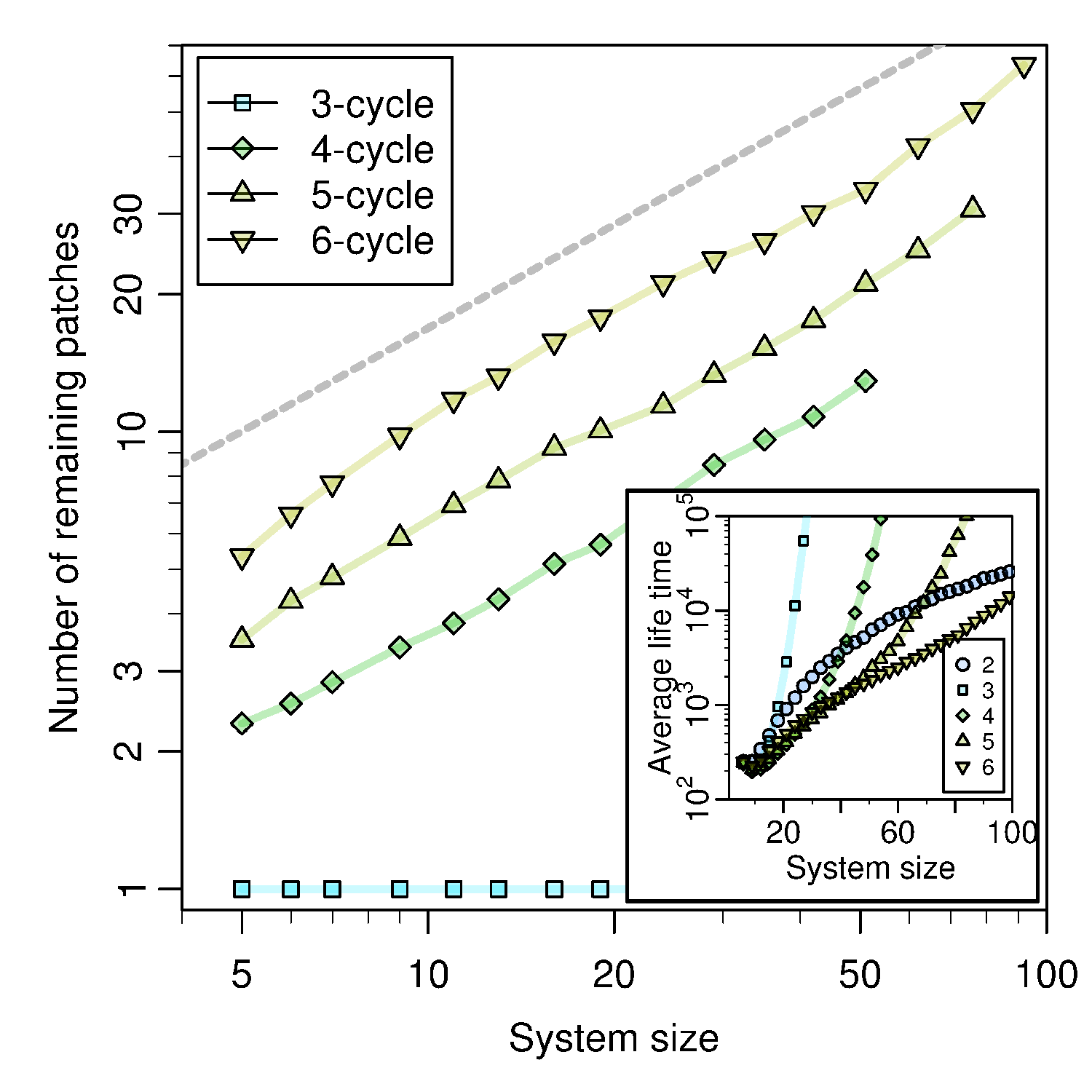}
\caption{ (Color online)
Investigation of breakdown of cycles: Average number of patches that are left when cycles of length $3,4,5$ and $6$ collapses due to fluctuations associated with the stochastic update of the system. The dashed line corresponds to the scaling $P\propto L^{0.75}$. The inset shows the average lifetime of the cycles of length  $2,\ldots, 6$ as function of system size. }\label{cyclelife}
\end{figure}

To study patch creation through cycles and their stochastic termination,  we show in Fig. \ref{cyclelife} a systematic investigation of the number of patches produced when a cyclic relationship is terminated due to the stochastic extinction of one species.  For a given cycle of length $C$, we initiate $C$ species with a cyclic relationship by randomly distributing the species in a square lattice of size $L \times L$. We let the system evolve until the dynamics spontaneously comes to a halt. Fig. \ref{cyclelife} shows the average number of patches left when a cycle is terminated as a function of system size $L$. It is clear that, even though the long cycles do not live as long as the short cycles (see inset of Fig. \ref{cyclelife}), they will create more patches when they terminate. Interestingly, the patchiness $P$ increases with the system size $L$ with an exponent close to $3 / 4$.

\subsection{Patch creation and biodiversity in the model}

\begin{figure}
\includegraphics[width=0.45\textwidth]{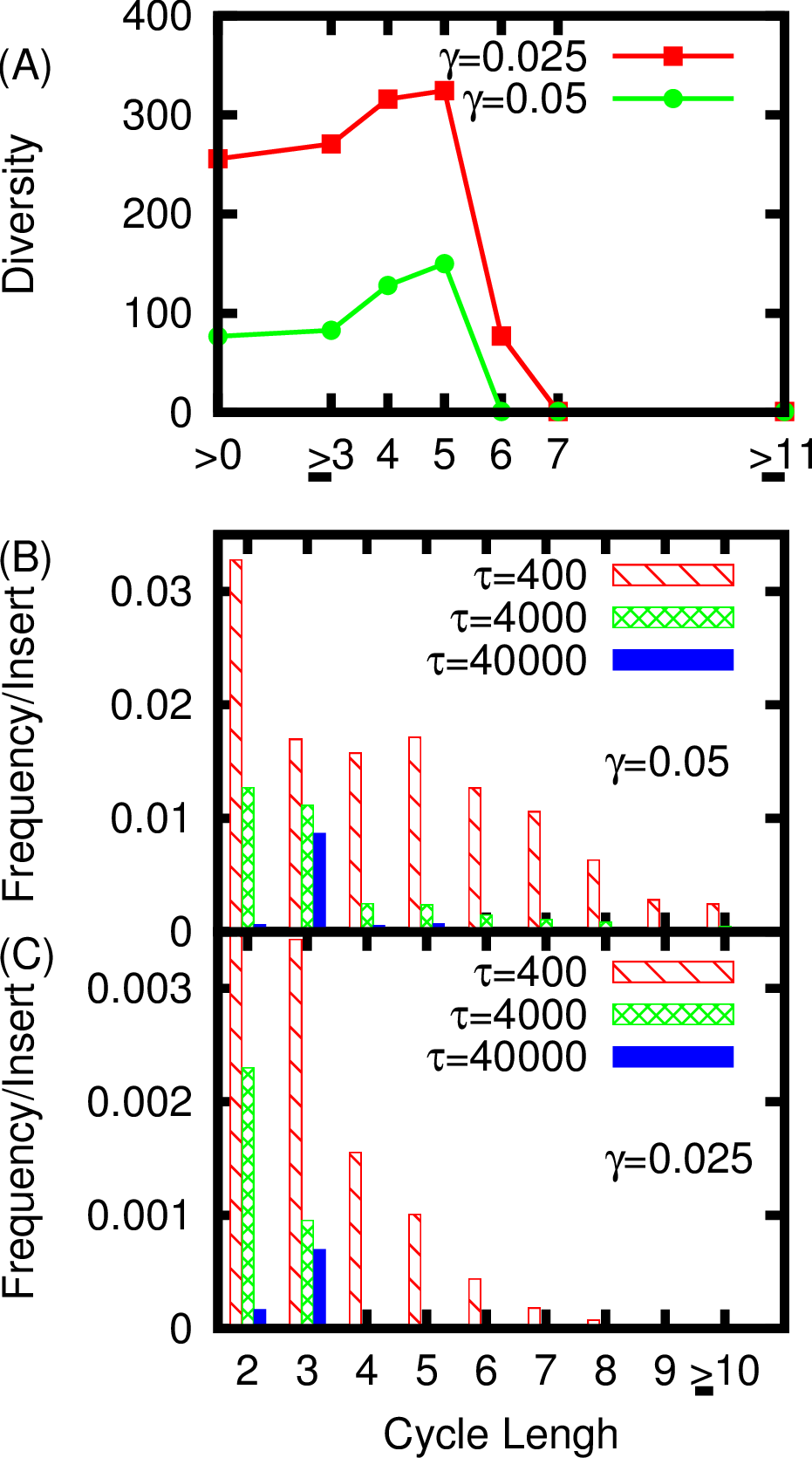}
\caption{ (Color online)
Quasistatic simulation of a system of size $L=200$. (A) Diversity of a steady state where short cycles of lengths below $2,3,4,\ldots$ are prevented from forming in the system. Removing cycles of length 2 is the same as allowing all cycles $\geq 3$.  { The connecting lines are for guiding the eye.}
(B, C) Probability of having an active cycle of a certain length at a time $\tau$ after the introduction of a new species for respectively (B) $\gamma=0.05$  and (C) $\gamma=0.025$. Notice that the probabilities do not add up to 1, which reflect the
fact that the majority of new species does not activate a cyclic relation. }\label{quasistatic}
\end{figure}

In order to examine which mechanism of patch creation is the most relevant for maintaining the high diversity, we run the quasistatic version of the model, where a new species is inserted into the system only after all the activity in the system has stopped. Occasionally, however,
there will be long periods where several species compete dynamically for the same area. { This is typically due to cycles of length 2 or 3, which have quite a long life time (Fig.~\ref{cyclelife} inset; see also  Fig.~\ref{quasistatic}BC). To shorten these periods, we perform the following procedure if the dynamics goes on for a time longer than $\tau_{limit}$: (i) When the active period reaches $\tau_{limit}$, randomly choose one of the active species (species that can invade its neighbor), $k$. (ii) Temporally eliminate all the outlinks for the species $k$ by setting $\Gamma(k,i)$ to zero for all existing species $i$. (iii) Run the invasion steps using the new $\Gamma$. (iv)With interval $\tau_{limit}$ repeat the procedure (i) to (iii) until all the dynamics are stopped.
After the system has frozen, the eliminated links are re-introduced (i.e., $\Gamma(k,i)$ will be set to the original values if species $k$ still survive somewhere in the system)  \cite{footnote1}, and the simulation is continued by introducing a new random species. In the following data, $\tau_{limit}$ is chosen to { be $4\times 10^4$}, but we confirmed { that $\tau_{limit}=4\times 10^3$} simulation did not change the average diversity (data not shown). Note that  the chosen value of $\tau_{limit}$ is  long enough for a linear relationship to come to a halt for system size $L=200$. }
The quasistatic simulation reflects the biologically interesting limit where 
the introduction of new species happens at a much slower than any ecological dynamics arising from species interaction and can, hence, be interpreted as resulting from speciation (rather than immigration).

In \cite{Mathiesen2011} we have confirmed that the quasistatic simulation has two metastable states when $\gamma<\gamma_c$, one with a high diversity and one absorbing state where $D=1$. The high-diversity state has a $D$-value close to the one for $\alpha=0.01$.

We performed quasistatic simulations with cycles of various degree. Simulation were started from a state of high diversity (taken from the steady state with $\alpha=0.025$ and $\gamma=0.05$), but with all interactions (i.e., elements in $\Gamma$ for initially existing species) set to zero at time $t=0$.  Subsequently, whenever a new species was introduced, the corresponding entries in the interaction matrix $\Gamma$ were determined according to the given value of $\gamma$, but if it would result in a cyclic relationship of length less than $ C_{min}$, the species was rejected, and a new species is introduced, which again was assigned random interactions according to $\gamma$. In this way, only cycles of length $C_{min}$ and above can be created.

The resulting diversity in the steady state for a system of size $L=200$ with $\gamma=0.05$ and 0.025 is shown in Fig. \ref{quasistatic}(A). The axis label $>0$ indicates $ C_{min}=0$, which is the original quasistatic simulation where all  cycles are allowed. The axis labels of, for example, “$\ge 3$” corresponds to $ C_{min}=3$, which allows all cycles of length $\ge 3$. Interestingly, removing cycles of length $2$, $3$, and $4$ even increases diversity, but further removal drastically reduces it. When cycles are of length 6 and larger, the system collapses to a single species ($D=1$) with $\gamma=0.05$,  while for $\gamma=0.025$ a diversity$>1$ is maintained. However, when even more cycles are removed, also the system with $\gamma=0.025$ collapses to the $D=1$ state. Accordingly, the cycles of length 5 and 6, in particular, crucially influence the transition point. This also indicates that patch creation by linear relationships, which should be present all the time, is not enough for maintaining high diversity.

The frequency of appearance of active cycles in the quasistatic simulation with $ C_{min}=0$ is investigated in Fig. \ref{quasistatic}(B) and (C), which shows the probability of having an active cycles of a certain length at a time $\tau$ after the introduction of a new species for (B) $\gamma=0.05$  and (C) $\gamma=0.025$, respectively. The length of the active cycle for a given moment is defined as the largest cycle within active species. We can see that, at { $\tau=400$}, there are many active cycles, but they vanish as time passes, and at { $\tau=40,000$}, most of the cycles left are of length 2 and 3, which do not leave patches. This again demonstrates that the long-lasting dynamics is not essential for keeping high diversity. Or, in other words, long lasting sustainable biodiversity is associated with patch creation occurring on a short timescale via the break-down of cyclic relationships.

{ It should also be noted that, even for { $\tau=400$}, the frequency of active cycle decreases with cycle length, as seen in Figs.~\ref{quasistatic} (B) and (C). Thereby cycles of length 4-6 become central for the creation of patches: Shorter cycles cannot create patches and longer cycles are too rare to make a difference.}

\section{Discussion}

We investigated the mechanism maintaining high biodiversity in a simple model ecosystem. High diversity is maintained if inter-species interaction parameter $\gamma$ is below a critical value $\gamma_c$. In the vicinity of $\gamma=\gamma_c$, the model shows bistability. One stable state is characterized by low diversity, which goes to $D=1$ as the rate of introduction of new species $\alpha$ approaches zero.  The other stable state is the high-diversity state, which exists even in the $\alpha \to 0$ limit. 
The transition from low- to high-diversity is triggered by a spontaneous increase of patchiness, while the collapse from high to low diversity is preceded by a decrease in patchiness. Thus the two transitions follow different paths:
\begin{itemize}
\item
{ The transition from low to high diversity is driven by the collapse of extended cyclic relations that create multiple patches of one species (metapopulations) within the range of another species. Each of these patches can subsequently serve as a seed or a shelter for newly arriving species. }
\item
{ The transition from high to low diversity is preceded by a reduction in patchiness. Low patchiness is equivalent to a lack of spatially separated shelters, leading to suppression of coexistence of antagonistic species.}
\end{itemize}

We further found that in order to maintain a balance between collapse and sustained patchiness, the system relies on very short-lived cyclic relationships involving more than five species, whose collapse creates a mosaic of mutually compatible species.

There still remain open questions in the transition observed in the model.
One question is the cut-off of the patch size,
which is independent of $\gamma$ and $L$ (Fig.\ref{1200sys});
we do not understand where the length scale comes from.
Another interesting observation is that,
in the histogram of diversity in Fig.~\ref{phasediagram},
the boundaries between the high and low diversity states are rather
$\gamma$ independent (around 15), even though the position of the peak in the
high-diversity state clearly changes with $\gamma$.
The $\gamma$-independent boundary
might suggest the existence of a critical minimal biodiversity $D^*$ needed to ensure the stability of a high-diversity state.
A minimal $D^*$ also translates into a
minimal system size of $L\approx 150$
to support the high-biodiversity state.

{ A natural extension of the model would be to include a death rate that create empty sites, which can be recolonized from neighboring sites (e.g. \cite{karlson}). This is equivalent to a small probability of species $u$ invading a site occupied by species $v$ even if $\Gamma(u,v)=0$.  Previously \cite{Mathiesen2011}, we tested the effect of death by emptying a fraction of sites prior to the introduction of new species in a quasistatic simulation. We found that the sharp transition to high diversity is maintained as long as this fraction is less than 10\%. With finite $\alpha$,  the transition is softened by a death rate that is small relative to $\alpha$. If the death rate is high compared to $\alpha$, diversity is expected to collapse. How exactly the transition between high- and low-diversity states depends on $\alpha$ and the death rate is a biologically relevant question, which we plan to investigate in the future.}

Overall, our model system supports biodiversity through a self-organized heterogeneity that is fed by the spontaneous collapse of cyclic relationships. This picture of emerging complexity contrasts the biodiversity associated with long-lasting dynamical cycles emphasized in the pioneering work of Buss and co-workers \cite{jackson,buss,karlson}. In this perspective the cycle duration shown in the insert of Fig.~\ref{cyclelife} is a minimal investigation of the Buss scenario, whereas biodiversity in our ``evolution'' scenario instead depends on the patches that are left when cycles collapse.

{\it Acknowledgement} This study was supported by the Danish National Research Foundation through the Center for Models of Life.

\begin{thebibliography}{10}
\bibitem{footnote1}
This can restart dynamics, if species $k$ left next to species $j$ with $\Gamma(k,j)=1$ and $\Gamma(j,k)=0$.
This is, however, unlikely if the long-lasting activity was due to an active
cycle.
\end{thebibliography}

\end{document}